\documentclass[fleqn,10pt]{wlscirep}
\usepackage[utf8]{inputenc}
\usepackage[T1]{fontenc}
\title{Seal and Sea lion Whiskers Detect Slips of Vortices Similar as Rats Sense Textures}

\author[1,*]{Muthukumar Muthuramalingam}
\author[1]{Christoph Br\"{u}cker}

\affil[1]{School of Mathematics, Computer Science and Engineering, City, University of London, London, UK}

\affil[*]{muthukumar.muthuramalingam@city.ac.uk}

\affil[+]{these authors contributed equally to this work}

\begin{abstract}
Pinnipeds like seals and sea lions use their whiskers in hunting their prey in dark and turbid conditions. There is no theoretical model or a hypothesis to explain the interaction of whiskers with hydrodynamic fish trails. The present work provides insight into the mechanism behind the Strouhal frequency identification from a Von-Karman vortex street behind bluff bodies, similar to the inverted hydrodynamic fish trail. Flow over 3D printed sea lion head with integrated whiskers of similar geometrical and material properties was investigated when being exposed to vortex streets behind cylindrical bluff bodies. It is found that the whiskers respond to the vortices by a jerky motion similar to the stick-slip response of rat whiskers on different surface textures. The Strouhal frequency of the upstream wake is clearly decoded with the time-derivative of the whisker response rather than the displacement response, which increases the sensing efficiency in noisy environments. It is hypothesized from the work that the time derivative of the bending moment of the whiskers is the best physical variable, which can be used as the input to the neural system of the pinnipeds.
\end{abstract}
\begin{document}

\flushbottom
\maketitle
%
%
\thispagestyle{empty}

\section*{Introduction}
Nature has provided sensing capabilities for different living organisms based on their living environment and the struggles, which they face in their habitats. For example, fish use their lateral line system to identify even the slightest movement in the water\cite{ref1}, crocodiles use miniature arrays of dome-shaped pressure sensors in order to identify the motion of its prey\cite{ref2}. Bats, whales, and dolphins use the bio-sonar system to locate and identify the objects\cite{ref3}$^,$\cite{ref4}. Similarly, land mammals such as rats and squirrels vibrate their whiskers, commonly known as `whisking’ to get information about their surroundings\cite{Knight2489}. Pinnipeds like Phocidae and Otariidae also possess highly sensitive whiskers which act as hydrodynamic receptors that can detect very small velocity fluctuations in the water which are caused by fish.

	Dehnhardt et al. demonstrated that Harbour seals respond to a dipole stimuli that ranged from 10 – 100~Hz, with a minimum perceivable water velocity of 0.25~mm/sec at 50~Hz\cite{Dehnhardt1998}. Later, Dehnhardt found that Harbour seals could track hydrodynamic trails generated by a model submarine even after a period of 25~sec, which corresponds to a chase length of 40~m\cite{Dehnhardt102}. Gl\"{a}ser et al. conducted the same experiment as Dehnhardt, however, with a California sea lion rather than the Harbour seal. They concluded that the California sea lion can also follow the wake trail, however, it could only track the object if the delay time was 7~sec or less\cite{Glaser2011}. California sea lions are even more sensitive than Harbour seals in dipole experiments. In the frequency range 20 to 30~Hz, they already responded to sinusoidal water movements that had a velocity amplitude of only 0.35~mm/sec\cite{Dehnhardt}. The differences in tracking ability between these two species is attributed to the different topology of the whiskers. Harbour seal whiskers have the shape of tapered beams with an undulating elliptic cross-section. This undulating geometry has been seen to reduce the self-induced `Vortex Induced Vibrations' (VIV)\cite{Hanke2665}. The VIV phenomena is caused by the alternating shedding of vortices from either side of the whisker cross section. In contrast, the California sea lion whisker has an elliptical cross-section which tapers toward the tip and has a constant perpendicular orientation over the length\cite{Hanke2013}. This geometry was found to not perform well in reducing the VIV. Miersch et al. compared the hydrodynamics of the isolated whiskers of the two species in rotating flow tank and quantified that the signal to noise ratio of Harbour seal whisker is ten times higher than that of the California sea lion whisker. Miersch et al. then attributed this large difference, to the reduced accuracy in wake tracking ability of the species\cite{Miersch3077} as previously seen by Glaser et al. They added that the self-induced oscillation of the sea lion whiskers, due to VIV, might act like a carrier signal which can be used in determining the self-swimming velocity. Beem and Triantafyllou investigated the flow over model whiskers, which were made out of plastic using stereolithography. When exposed to vortex streets originating from an upstream cylinder, they described the whisker motion as ‘slaloming’ with the vortex street. The amplitude of vibration was very large in the wake of the cylinder when compared with the clean flow conditions (named in the following as Wake-Induced Vibrations (WIV) to distinguish from the self-induced VIV). This increased amplitude of the whisker vibration is understood to be the reason for the increased sensitivity of Harbor seals \cite{beem_triantafyllou_2015}. Schulte-Pelkum demonstrated that Harbor seals can effectively follow biogenic trails of same species. This tracking ability could be used by the seal pups to follow lactating mother seals during feeding time\cite{Schulte-Pelkum781}. In this study, the path of the trail-following seal was congruent to the path of the pilot seal for most trials. However, in some of the experiments the follower crossed the path in a zig-zag manner to redefine its course. Additionally, it was inferred that the trail-following seal was able to resolve the inner structure of the wake while tracking. \\

\begin{figure}[h!]
\centering
\includegraphics[width=0.9\linewidth]{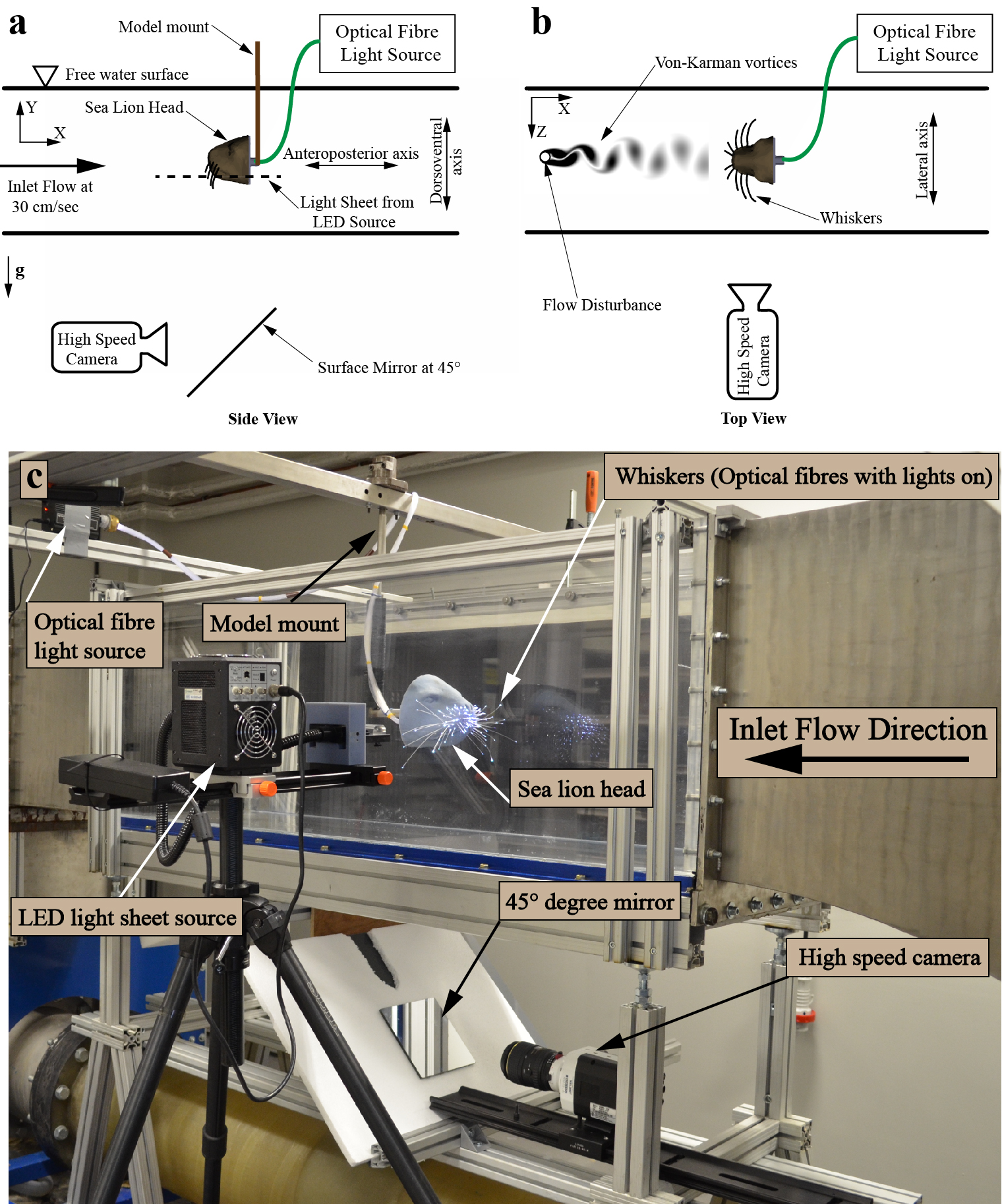}
\caption{a) Schematic diagram of the experimental set-up (side view) b) Schematic diagram of the experimental setup (top view). Von-Karman vortex street is shown for representation c) Actual experimental set-up (Note that the tunnel is not filled with water to get a clear view of the model)}
\label{fig:Experimental_Setup}
\end{figure}

    Wieskotten et al. found that by moving an artificial flexible fin, in order to produce a wake trail, the Harbour seal could indicate and follow the direction of the moving fin up to a delay of 35~sec \cite{Wieskotten2194}.As such, it was concluded that the seal could have used the information in the high velocity region between the two counter-rotating vortices and the structure of the vortex street as a hydrodynamic cue to track the direction. In a successive experiment, Wieskotten et al. found that the Harbour seal could discriminate the size and shape of the objects by tracking their corresponding wakes \cite{Wieskotten1922}. However, they also concluded that the size and shape discrimination ability relies on more than one hydrodynamic parameter. Kr\"{u}ger et al. argued that Harbour seals use vortex rings as the primary source of directional discrimination \cite{Kruger}. By simulating a variety of test cases with single vortex rings, it was concluded that the minimum perceivable angle between the anteroposterior axis and the path of the vortex ring was less than 5.7$^\text{o}$, however, the minimum value was not found because of technical limitations. Niesterok et al. experimented on harbour seals by imitating artificial flatfish breathing currents. This was done by producing intermittent jets from an underwater surface and found that the seals could identify velocities in the range of 2-2.5 cm/sec\cite{Niesterok2364}. This value is significantly higher than that of the value (0.25~mm/sec), attained from dipole experiments\cite{Dehnhardt1998}. The self-induced velocity variations because of the animal’s own movement could be the reason behind this reduced sensitivity. Murphy et al. tagged an accelerometer on a supraorbital whisker of a real Harbour seal and concluded that the seal whiskers vibrate over a broad range of frequencies in an upstream hydrodynamic disturbance condition \cite{Murphy2017}. \\

In biomimetic applications, sensors based on mammalian whiskers were developed and successfully tested. Subramaniam et al. tested a 3D printed individual Harbour seal whisker in a towing tank and quoted that the VIV of Harbour seal whisker is lesser when compared with cylindrical whisker\cite{Subramaniam2017}. Solomon and Hartmann used conical beam whiskers to trace the complex surface shapes and reconstructed it using bending beam theory \cite{Solomon2006}. They used multiple tandem whiskers to get the velocity profile shape in a free jet flow. Wake tracking ability of multiple whiskers was demonstrated by Eberhardt et al which was inspired from the seal’s whisker array \cite{William}. They used eight cylindrical whiskers on a torpedo like body to detect the wake interference from an upstream model submarine. They concluded that multi-point measurements using whiskers like sensor will result in accurate tracking of wake flows and can be used in flow sensing. \\ 

	Even after all of the previous work, in studying pinniped species and their hydrodynamic trail tracking abilities, there are still some fundamental questions which remains unanswered. For example, what effect does the animal’s head have in altering the strength of the signal, how does the arrangement of whiskers with respect to the vortex direction play an effect, do they make use of multiple whiskers and how do they interact with the alternating type of oncoming vortices similar to wakes produced by fish. Fish swim within a very narrow Strouhal number band, St $\simeq$ 0.2 and the swimming speed of the fish is directly proportional with its tail beating frequency and its length\cite{Nudds2244}$^,$\cite{BAINBRIDGE109}. Hence, the frequency is the single most important feature to identify the swimming speed of the fish. The wake behind a swimming goldfish, where the length of the fish was 10~cm, produced a wake trail of measurable velocities until a time of 30~sec passed\cite{Hanke1193} and this time lapse should theoretically scale with the length-scale of the fish. Since Harbour seals feed on fish with an average length of 30~cm \cite{RJ2009} the information left behind the wake of the fish is certainly a strong hydrodynamic indicator for the predators. The important question here is, how this information is best decoded in an otherwise noisy environment by the seal’s neural system . From a fluid dynamics point of view, when the seal swims with its whiskers protracted forward, the fluid-induced drag forces cause the whiskers to bend in the direction of the mean flow and the swimming speed is directly proportional with the deflection of the whisker.  Since whiskers do not have nerves \cite{Nerve}, it is thus the nerve endings in the whisker pads that read the information about the bending moment at the base of the whiskers. These nerves are connected with the somatosensory system in the brain of sea lions\cite{Sawyer}. Hence, a wake trail of a fish that swims upstream should change the bending moment of the whisker depending on the velocity distribution in the wake and this could be encoded in the neural system. In literature, a satisfactory answer for this question could not be found because of the following reasons: first, it is very hard to carry out an experiment with living seals or sea lion with a controlled environment such as constant swimming speed without any lateral head movements, which is likely to cause a change in whisker motion; second, it is very hard to track the whiskers movement and the vortex convection in the same experiment simultaneously when a wake generator is moved upstream to the animal. This is the primary objective of this work and it is the first kind of experiment on pinniped whisker-like structures, which is almost similar to the realistic situation.\\
    
      The experiments were conducted in return type open surface water tunnel with a velocity range of 0.1 to 2~m/sec. Flow straighteners were installed before the convergent section in the settling chamber to  get uniform flow in the test section. The test section is 40~cm wide x 50~cm depth x 120~cm in length and transparent in all the sides to provide an optical access for flow studies. All the experiments were conducted at a flow velocity of 30~cm/sec based on previous studies \cite{Miersch3077}. The model was placed at 70~cm from the inlet at the centre of the test section as shown in Fig. \ref{fig:Experimental_Setup}. The experiments were split into two stages. The first stage was to track the motion of the whiskers, by focusing the `Eye Motion' camera to just one side of the sea lion model, as shown in Fig. \ref{fig:Experimental_Setup}b. And then a subsequent motion tracking experiment was carried out with the `Eye Motion' camera underneath the tunnel (Fig.  \ref{fig:Experimental_Setup}a) in order to observe all of the whiskers concurrently. The final experiment was a particle image velocimetry (PIV) study, in order to see how the vortical structures shed from a flow disturbance (Fig. \ref{fig:Experimental_Setup}b) interact with the whiskers. It should be noted here that for the motion tracking measurements, the LED light sheet source used for the PIV is removed and the `Eye Motion' camera is installed at the same location. \\
    
\section*{Results}

\begin{figure}[!ht]
\centering
\includegraphics[width=0.71\linewidth]{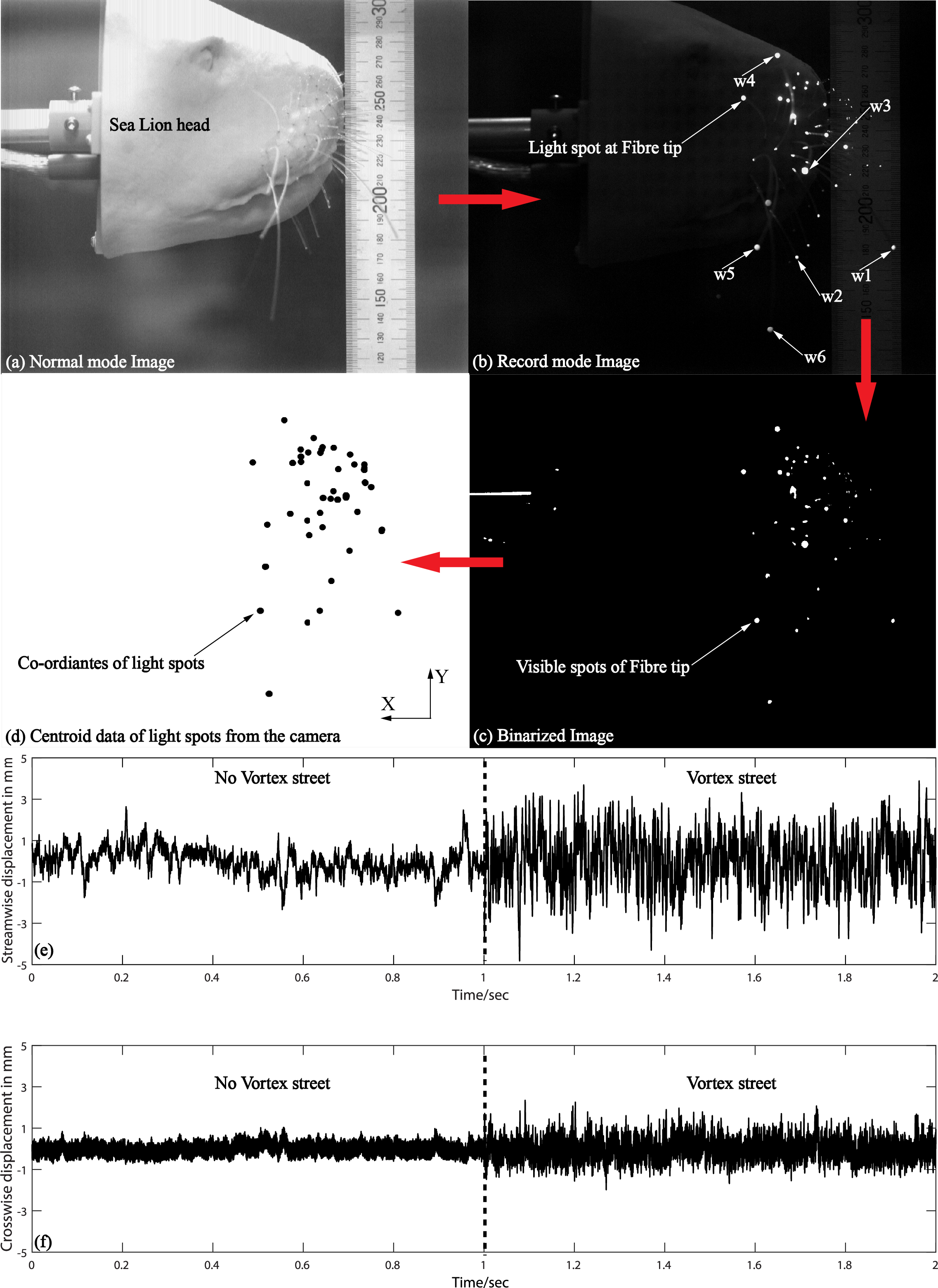}
\caption{a) Normal mode photograph of the model from the camera (Scale in mm) b) Image from the camera with reduced light and fibre optics on (the tips of fibres are visible as light spot) Note that the whisker numbering is used in the discussion of results c) Binarized image in which the fibre tip light is converted into white spots greater than certain threshold d) Coordinates from the light spots in x-y plane.  e and f) Displacement of whisker tip in  streamwise (anteroposterior) and crosswise (dorsoventral) direction with and without vortex street.}
\label{fig:binarisation}
\end{figure}

The results of six whiskers were used in this study and they are shown in Fig. \ref{fig:binarisation}b. The whiskers could be identified as w1, w2, w3, w4, w5 and w6 in which the whisker w2 is about 50~mm length. All other whiskers were about 100~mm in length. The tip motion of the whiskers were tracked in a plane which is parallel to the X-Y plane as shown in Fig. \ref{fig:Experimental_Setup}a. The time history of each whisker can be extracted from the 'Eye motion' camera as a coordinate file and  the data for whisker w3 is shown in Fig. \ref{fig:binarisation}e and f. The x-direction of the whisker bending is along the anteroposterior body axis aligned with the flow direction in the tunnel while the y-direction is along the dorsoventral axis. It should be noted here that when the whisker experiences an upstream wake the displacement in the streamwise direction (anteroposterior direction) is changed significantly due to WIV when compared with the crosswise displacement (dorsoventral direction)  which occurs because of the VIV. As explained before, the deflection of the tip is a function of the bending moment caused by the corresponding fluid forces in that respective direction. Hence the spectrum from the whisker tip data along the x axis corresponds to the bending spectra and the movements along y axis corresponds to the VIV spectra.

\begin{figure}[h!]
\centering
\includegraphics[width=0.9\linewidth]{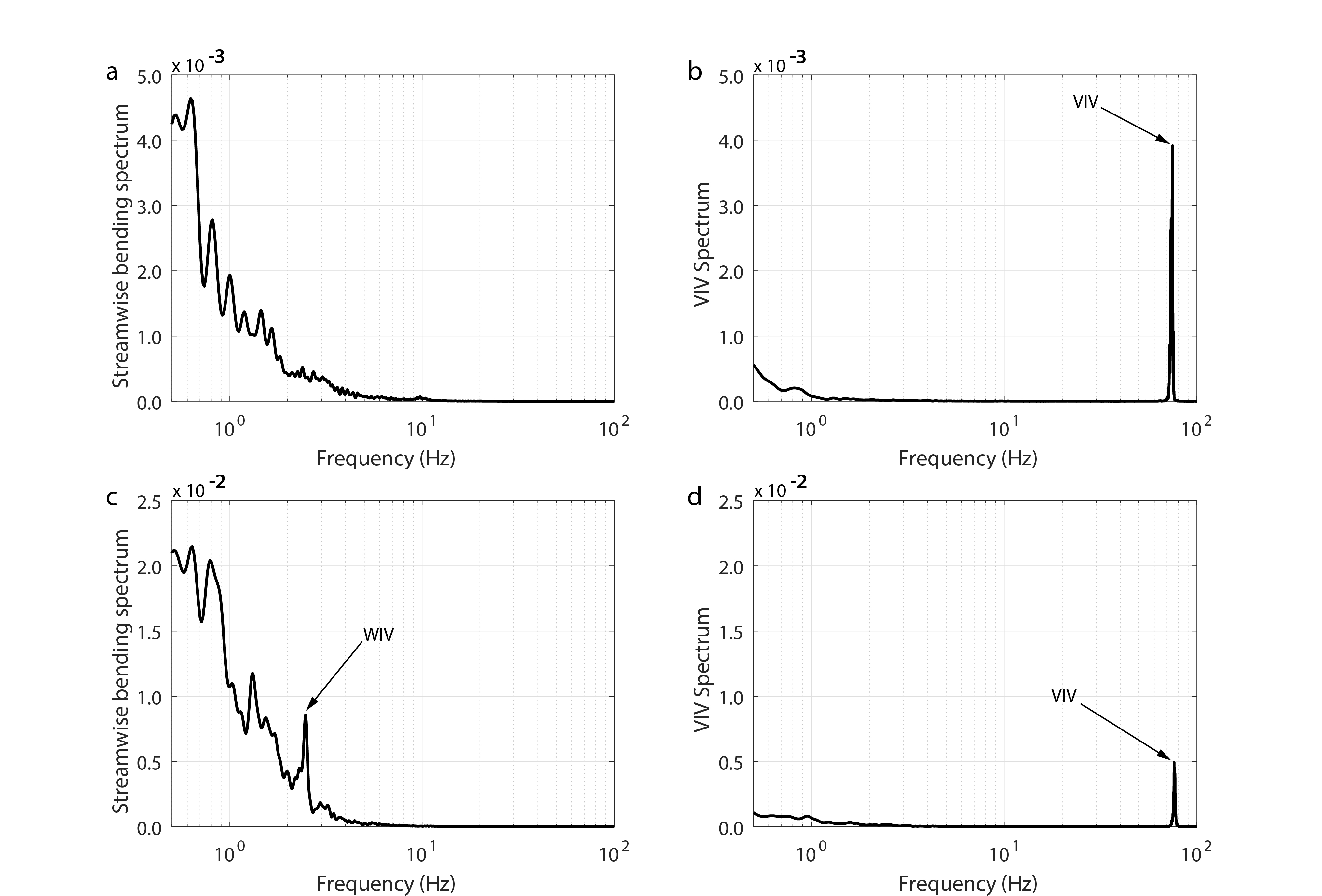}
\caption{Spectrum of whisker tip fluctuations (whisker3 w3) in undisturbed water flow (a,b) and when exposed to a vortex street (c,d). The VIV spectrum is highlighted in the motion along the dorsoventral axis (distinct peak at 75~Hz) while the WIV (Streamwise bending spectrum) show up in those along the anteroposterior axis corresponding to the swimming direction (distinct peak at 2.4~Hz)}
\label{fig:freestreamandvortexcomparison}
\end{figure} 

Figure. \ref{fig:freestreamandvortexcomparison}a shows the spectrum of the movement of whisker w3, which is almost aligned with the lateral axis perpendicular to the sea lion head. The spectrum shows frequencies below 2~Hz for the whisker movements in flow direction (vibrations along the anteroposterior axis) (Fig. \ref{fig:freestreamandvortexcomparison}a) and a strong peak at 75~Hz in Fig. \ref{fig:freestreamandvortexcomparison}b (vibrations along the dorsoventral axis). The peak at 75~Hz reflects the VIV caused by the alternating shedding of shear layers from the whisker \cite{VIV}. The Strouhal number based on the flow speed of 30~cm/s, diameter of the whisker (0.75~mm) and the spectral peak at 75~Hz  is 0.19. The corresponding Reynolds number, where the whisker diameter is the length scalar, is 225. The observed Strouhal frequency is in good agreement with previous literature at the given Reynolds number\cite{Cylstrouhal}. It is important to note that the 75~Hz component of the spectrum is the dominating mode of the whisker vibration in undisturbed flow (swimming in quiescent water). When the same whisker is facing a vortex street, which is generated by placing a 25~mm diameter cylinder upstream of the model, there are considerable changes to its response. Figure. \ref{fig:freestreamandvortexcomparison}c demonstrates the WIV response in the flow direction with a distinct local peak at 2.4~Hz. The Strouhal number based on the spectral peak of 2.4~Hz and an upstream cylinder with a diameter of 25~mm is 0.2. This corresponds exactly to the value cited in the literature for a cylinder Reynolds number of 7500 \cite{Strouhalref}. At the same time, we observe energy at lower frequencies < 2~Hz.  The first subharmonic of the WIV is seen as a peak at 1.2~Hz. Furthermore, some energy is contained in a wavy motion lower than 1~Hz which is thought to be resulting from a low-frequency sloshing motion in the open-surface water tunnel. The VIV frequency of the whisker is still present in Fig. \ref{fig:freestreamandvortexcomparison}d as a distinct peak at 75~Hz. However, the magnitude of this peak is now at least twice less than the WIV frequency. It should be emphasized here that sea lion whiskers with smooth elliptical cross-sections have shown to reduce VIV magnitudes when compared to artificial whiskers with a cylindrical body. The presented results, with cylindrical fibres, suggest that even when such simplified whiskers experience a wake flow, the WIV response is still of greater importance than the VIV response. Therefore it can be concluded that the WIV response in the natural situation is even more significant.  \\

\begin{figure}[h!]
\centering
\includegraphics[width=0.76\linewidth]{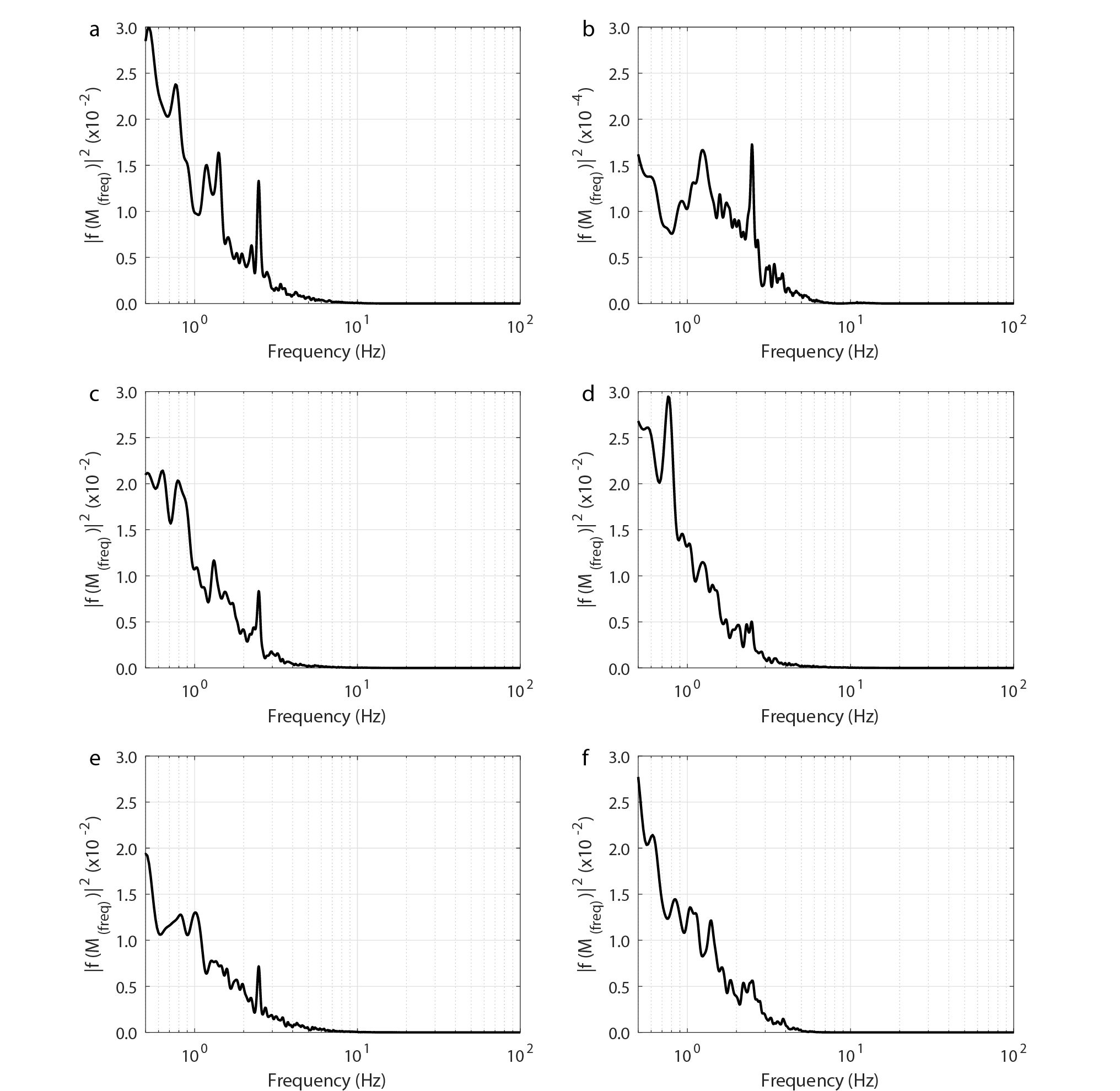}
\caption{Spectrum of tip fluctuations of different whiskers along the anteroposterior axis when exposed to a vortex street of a 25mm cylinder in front (a: w1, b: w2, c: w3, d: w4, e: w5, f: w6). All whiskers are of length l\textasciitilde100mm, except whisker w2 is only of medium length with l\textasciitilde54mm, therefore note the different axis scale in plot b.}
\label{fig:cylspec}
\end{figure}

Note, that similar observations of the relative importance of WIV against VIV were made for the other whiskers as well. However, their individual WIV response depends largely on the location, orientation, and length of the whisker along the head, as seen in  Fig. \ref{fig:cylspec}a, c and e show the previously described  distinct WIV peak at 2.4~Hz. However, this peak is not always dominant when compared with the range of frequencies <1~Hz. Furthermore, there is no indication of a strong peak at all in Fig. \ref{fig:cylspec}d and f. Finally, the medium length whisker exhibits this peak only when magnifying the displayed axis in Fig. \ref{fig:cylspec}b about a factor of 100 against the other plots. This illustrates that, by purely using the bending moment as the signal for further neural processing might not be sufficient enough to distinguish between noise and vortex streets. It could result in an unsuccessful attempt to track the vortex source. It should be noted here that by comparing the left and right whisker pair the signal strength of Strouhal frequency can be amplified\cite{Muthu}. This result is similar to the adaptive filter which cancels the self-induced noise in the lateral line of fish\cite{MONTGOMERY1994145}.\\
 
\begin{figure}[h!]
\centering
\includegraphics[width=0.8\linewidth]{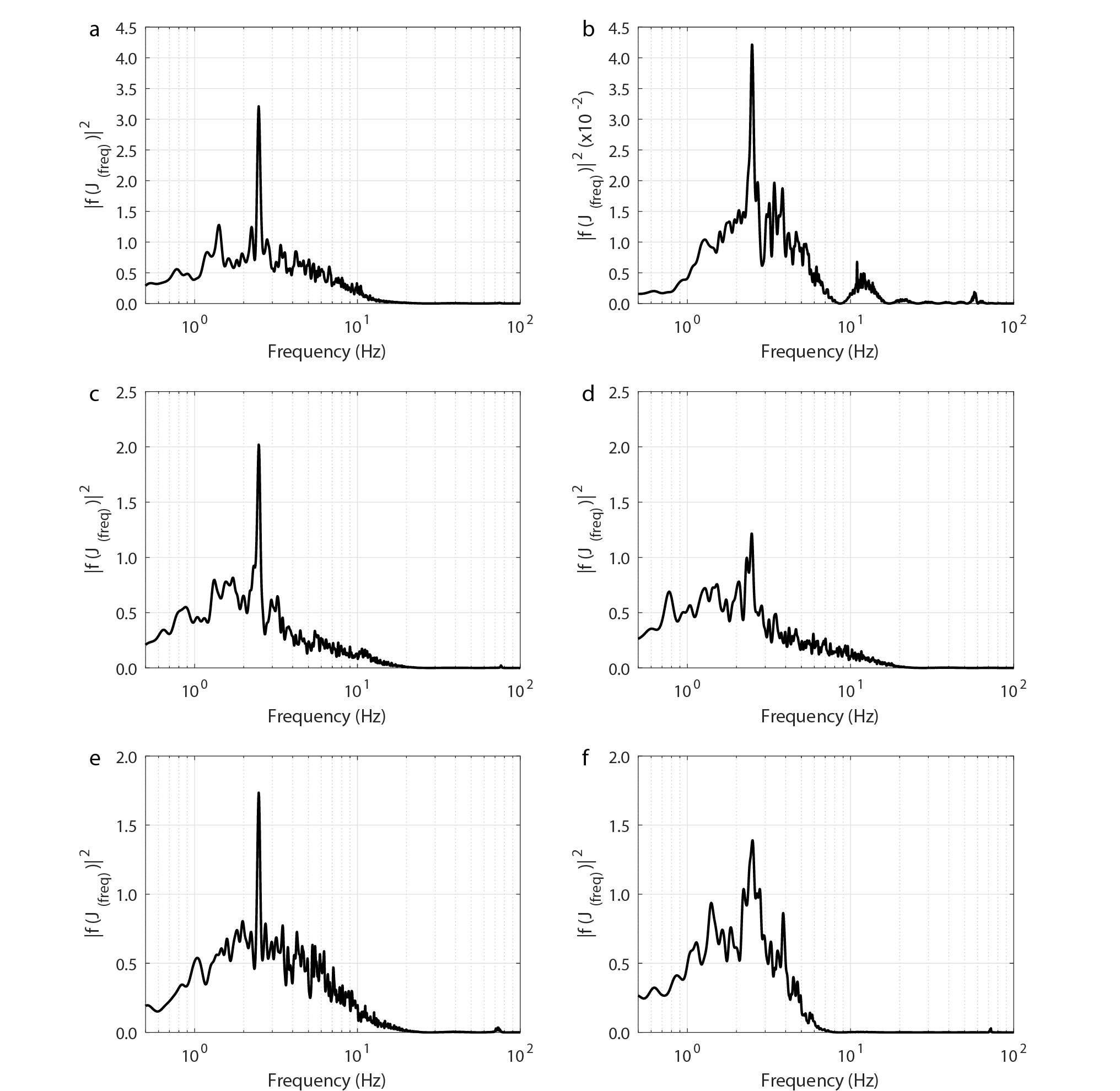}
\caption{Jerk spectrum of different whiskers along the anteroposterior axis when exposed to a vortex street of a 25mm cylinder in front (a: w1, b: w2, c: w3, d: w4, e: w5, f: w6). Note the peak at 2.4Hz in all plots }
\label{fig:cyljerk}
\end{figure}

In Fig. \ref{fig:cyljerk}, the spectrum of the time-derivative of the tip-displacement signal is plotted for all six of the tracked whiskers. It can be seen that for all whiskers a dominant peak is highlighted at 2.4~Hz, which is the Strouhal frequency of the cylinder. Earlier it was argued that the whisker’s tip movement over time reflects the temporal change in the bending moment acting at the whisker base. Hence, following the same argument, the time-derivative of the motion signal is the function of the ‘jerk’ at the whisker base. The jerk spectrum can then be seen to clearly increase the visibility of the dominant frequency, when compared to that of the bending moment spectrum. As such, this is clearly a very good method to identify the frequency of the shed vortices. If the neural cells respond to the jerk signals then they trigger pulses from all whiskers at the same rate and for brain cells, it is one common frequency which needs to be selected out.  \\

\begin{figure}[h!]
\centering
\includegraphics[width=0.825\linewidth]{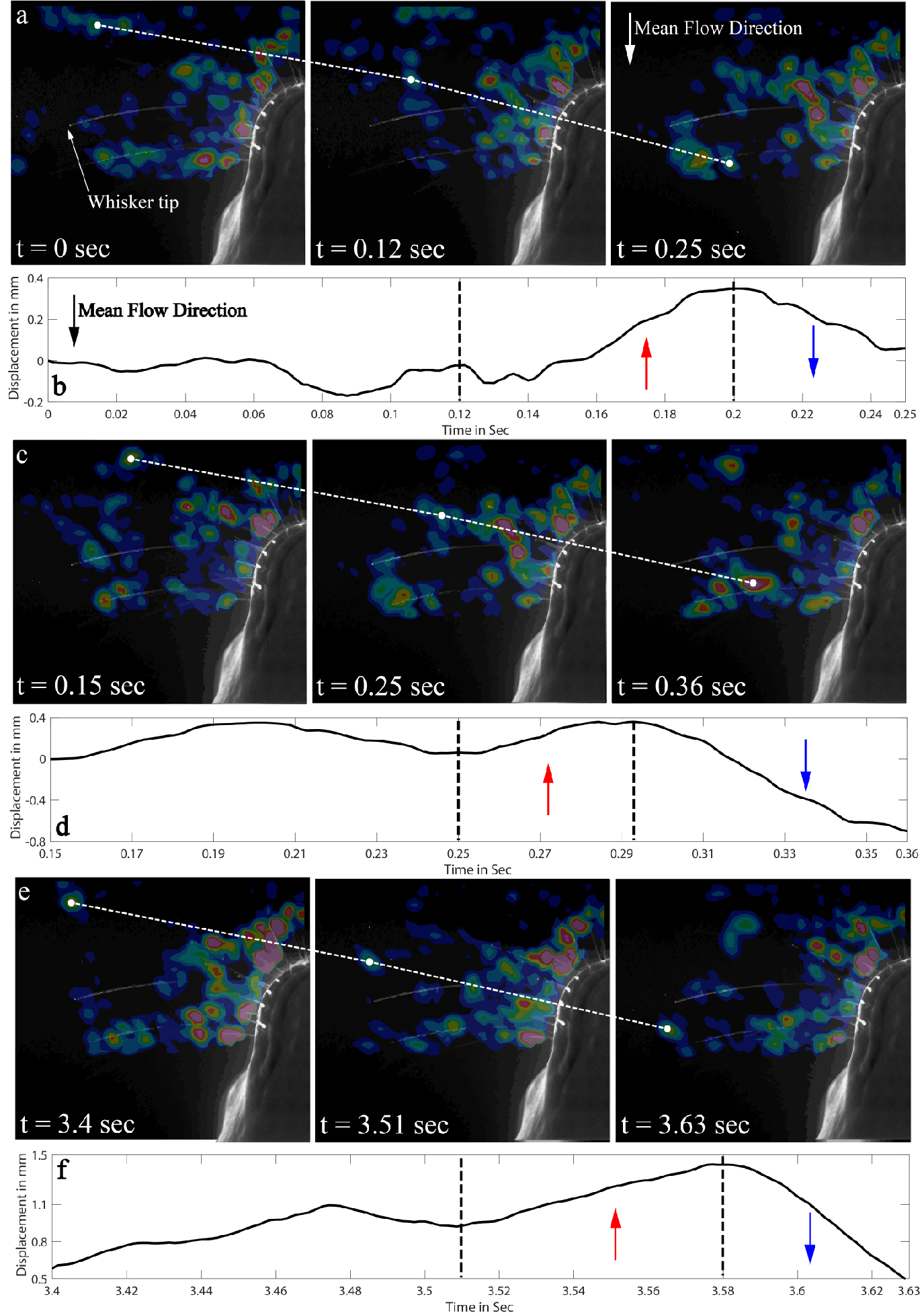}
\caption{a,c,e represents the Vortex core motion on three time instants. Mean flow direction is indicated in the (t=0.25 sec of a) Vortex cores are marked with filled white circles. b,d,f represents the Motion of whisker w3 on time scales corresponding to the vortex core motion in a,c and e. Note: The blue and red arrows in b,d and f indicates the motion of the whisker in flow direction and against flow direction respectively.}
\label{fig:vortex}
\end{figure}

\begin{figure}[h!]
\centering
\includegraphics[width=0.9\linewidth]{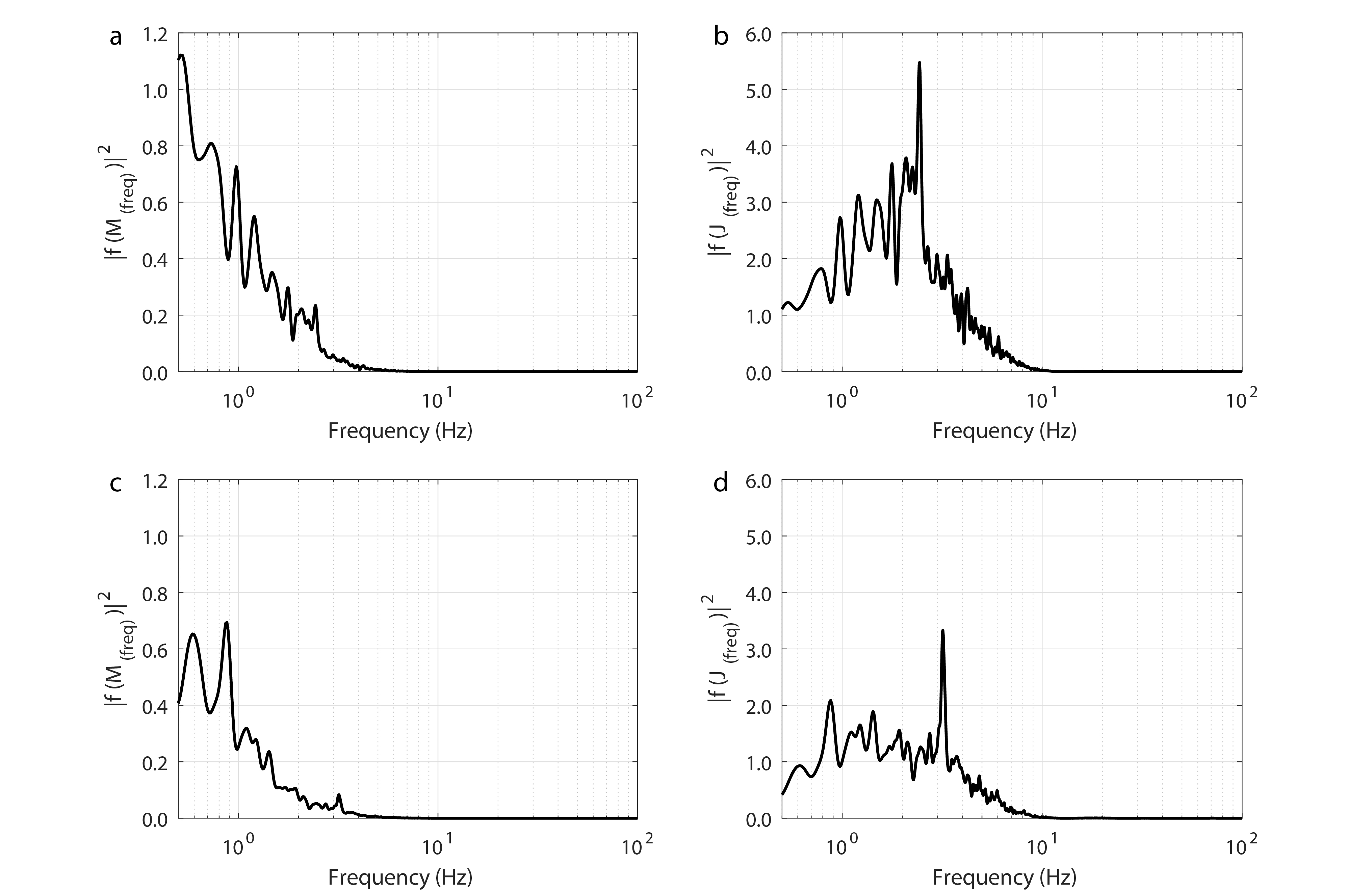}
\caption{a and c  bending moment spectrum of 25mm and 19mm cylinder wake. b and d are jerk spectrum of 25mm and 19mm cylinder wake. Note that all plots are for tip fluctuations of Whisker w1 in anteroposterior axis }
\label{fig:25mmand19mmcylcomparison}
\end{figure}

\section*{Discussion}

To better understand the coupling between local flow dynamics and the induced whisker motion, the experiments were repeated with simultaneous PIV flow field measurements and whisker tip recordings. In these experiments, the 25mm cylinder located upstream of the body was placed at a lateral offset from the body axis (4 Diameters, i.e. 100~mm offset in ‘z’ direction) such that the vortex street faces the whiskers only on one side of the model. The same side as the LED light-sheet illuminated the flow. PIV was used to study the vortex interaction on the local whisker elements. The vortex rollers shedding from the upstream cylinder are identified as regions of concentrated Q-values \cite{jeong_hussain_1995},  which are sometimes difficult to be seen as the presence of short whiskers and the light reflections from the model head affect the image quality. In the following experiments, images were selected where the location of the shed vortices could clearly be followed over time while passing whisker w3. In Fig. \ref{fig:vortex}a, from time 0 to 0.12 sec a shed vortex from upstream approaches the whisker w3 at a whisker length of about 75\% and initially, no considerable movement of the whisker except minor scale oscillations are seen (see Fig. \ref{fig:vortex}b). However, when the vortex gets close to the whisker, the whisker tip starts to move towards the vortex (direction anterior). It means that the whisker bending due to the mean flow drag is relaxing a bit due to the presence of the vortex next to the whisker tip. As the vortex passes over the whisker (as shown in time t=0.25 sec  of Fig. \ref{fig:vortex}a), the whisker tip reverses its motion direction and finally trails the vortex downstream (direction posterior) until it reaches its initial equilibrium position. In Fig. \ref{fig:vortex}c at time t=0.25sec a successive shed vortex from the cylinder has approached the same whisker. When getting closer,  the whisker tip again moves first towards the vortex against the incoming flow (direction anterior) and then reverses the direction when the vortex passes over the whisker. In Fig. \ref{fig:vortex}d at time t=3.4sec the path of another shed vortex is crossing the whisker tip directly. A similar forward-backwards motion pattern as described above is seen again.  The observed motion pattern can only be explained by the presence of pressure gradients in the vortex cores, as the sense of rotation of the vortices does not make a difference in the observed whisker behavior. If the sense of rotation played a role, the velocity variation along the whisker would cause the whisker tip to move always with the highest streamwise velocity inducing the largest drag along the whisker, which is sometimes on the left and sometimes on the right of the position where the vortex is passing the whisker.  In addition, it would not explain the relaxing behaviour once a vortex core is getting close to the front of a whisker. Instead, it is the pressure-gradient which is responsible for this motion. A vortex core in water flows is a region of low pressure when compared to the ambient flow because of the conservation of angular momentum \cite{Green1995}. When a vortex, being shed from the cylinder and embedded in the mean flow is approaching the whisker, the low-pressure region in the vortex core causes the whiskers to move first towards the vortex center, by reducing the bending moment on the whisker which is induced by the drag force of the mean flow. On the other hand, when the vortex passes the whisker, the whisker bending is increased again as the low-pressure core is dragging the whisker in addition to the mean flow drag. These small-scale movements associated with the time are repeated each time when a vortex core is moving across the whisker, and these events are representative for a  measure of the frequency of a vortex street in a flow. In the displacement domain, these events are small and periodic but not sufficiently strong to be detected against the noise floor or other events in the flow. Nevertheless, when we differentiate the signal in time, these small-scale associated repeated phenomena are amplified in time. In order to study the power of this signal processing strategy,  another experiment was carried out to investigate the response of whisker w1  where the diameter of the upstream cylinder was chosen smaller with only 19mm. This reduced cylinder size provides a considerably smaller signal strength of the vortex street, as the size and the vorticity in the vortex cores scales with the diameter of the bluff body. A minor local peak is observed in Fig. \ref{fig:25mmand19mmcylcomparison}c at 3.1 Hz for the 16mm cylinder which corresponds to a Strouhal number of 0.2. This peak is dominant in the jerk spectrum as shown in Fig. \ref{fig:25mmand19mmcylcomparison}d,  similar to the Strouhal peak for the 25mm cylinder shown in Fig. \ref{fig:25mmand19mmcylcomparison}b. In conclusion, even the vortex street at low signal-to-noise-ratio (SNR) leaves a very distinct peak in the jerk spectrum. The SNR  for the 19mm cylinder calculated in the range between 1 to 5 Hz by using the bending moment spectrum is 1.4, and the SNR by jerk spectrum is 3.9. It means there is a signal amplification by a factor of 3 by using jerk spectrum. It bolsters the hypothesis that the jerk is the better indicator of vortex streets.\\

A recent study on rat whiskers revealed that the best way for texture coding in rats is through the stick-slip movements which are discrete short lasting kinematic events \cite{Oladazimi2018}. These events are visible when the signal is differentiated in time. Similarly,  for sea lion whiskers, the Strouhal peak gets more evident when we use the ‘jerk’ spectrum which resulted in hypothesizing that there could be an analogy in the neural signal processing strategy of the whisker sensing in the different species. Coming back to the whisker mechanics, we will redefine the way of interpreting the interaction of vortices on whiskers. First, when the seal or sea lion swims, because of the fluid forces the whisker bends with the flow, this is comparable to the stick phase in rat whisker, because of the friction between the whisker and the surface it sticks on the surface. Secondly, in rats, the slip like events are when the whisker moves towards the whisking direction more than the amount of when whisking in the air. Accordingly, when a seal swims through a vortex core and feels them on their whiskers, they move forward against the swimming direction or ‘slips’ from the bending direction. However, here comes a slight difference in whisking mechanisms, once a rat whisker moves beyond an area of a surface there is no communication on the whisker motion from the trailing surface. On the other hand, when a seal whisker moves through the vortex core its effect is felt even on the downstream as the additional force effect which causes the whisker to respond fast in the stick phase. It is crucial to add the fact that whiskers are tactile sensors, rats and pinnipeds use them for identifying the surroundings. For a rat, the primary information is about the surface texture which provides them accurate representation and topology of the surrounding. For pinnipeds like seals or sea lions, the vortices are the primary source of hydrodynamic signatures of preys, which is similar to the textures. Hence, the short kinematic events like stick-slip mechanisms are common for both the species even though the behavioural and the evolutionary processes are different.

To summarize, this is the first work to study the whisker mechanics mounted on a sea lion head similar to realistic situations. The method of using multiple fibre optic cables of similar mechanical properties as whiskers was used to track the whisker motion for different cases. It was done without interfering the flow or without using force transducers that increases the complexity. Furthermore, with the use of the eye motion camera, the collective response of all whiskers in the field of view can be tracked successfully. We have identified that the WIV response in the lower frequency range is at least one order more intense than the VIV response of the whiskers. It is claimed here that the usage of elliptical cross-section whiskers reduces the VIV response and amplifies the SNR. For smaller wakes (less intense vortices) the time derivative of the whiskers response provides a clear indication of vortex frequency. From PIV results, it is shown that the whiskers slip from the bending direction when it encounters a vortex, and when it passes over the whiskers it sticks with a higher rate than the slip phase. This small jerky movement, which occurs in a shorter time scale, enlarges in the derivative domain, which amplifies the Strouhal frequency. From the results, it is concluded that the stick-slip events are common for the whisker sensing and it is the informative part from the whiskers which neural cells can encode and discriminate between different vortex frequencies.
 \\
 
\section*{Methods}

\subsection*{Model and Whisker setup}
Scanned data of sea lion body was acquired from Museo delle Scienze, Italy. The digitized data was cleaned and converted into a 3D model using CATIA. On the head of the sea lion, 38 holes of 0.8mm diameter were modelled on either side of the head on six rows as mentioned in Sawyer et al\cite{Sawyer}. The model was 3d printed at the City University of London. Fiber optic cables of 0.75mm diameter (based on Summarell et al)\cite{diameter} was inserted through the holes from the back end and was cut into the required length of the whiskers which ranges from 5 mm to 115mm as mentioned by Sawyer et al\cite{Sawyer}. All the fibers were glued at its surface to provide a fixed cantilever condition, and the open ends of the fibres were bunched and cut to form a flat surface to illuminate with a light source outside the water tunnel. Hence, the fiber optic cable acts as a whisker and the movement of whisker can be tracked with its light at the tip of the fiber. The optical fibres were made of polymethylmethacrylate (PMMA) plastic material which has an elastic modulus value of around 3.5 GPa (from Leal-Junior et al\cite{LEALJUNIOR2018106}), and the value of the seal whisker is in the order of 4 GPa as mentioned by Subramaniam et al\cite{Subramaniam2017}. Since both the geometrical and the mechanical properties of the whiskers are comparable, we expect this model to act as a passive whisker successfully. However, there are some limitations in this study, the fiber is circular in shape when compared to an elliptical section of the sea lion whisker. The diameter is constant for all fibers, and it is constant along the length, however for sea lions the whisker diameter changes throughout the length and the diameter changes for different whiskers. \\

\subsection*{Camera Setup and data processing}
A high-speed camera (ProcImage 500-Eagle high-speed camera) at a sampling rate of 250 Hz was used to track the motion of the whiskers. It had a pixel size of 1280 x 1024 which relates to a physical dimension of 238 x 190 mm. The built-in function in the camera calculates the centroid of a light spot automatically using the barycentre mode. The camera function is explained in Fig. \ref{fig:binarisation}a-d. When the surrounding lights are off, and the fibre optic lights are switched on, the camera records the images and automatically does the image binarization to discriminate between the light tips and the surrounding. Then it calculates in real-time the centroid of the light spots in pixels for all the visible fibre tips and saves those as an array of the X-Y coordinate data file. The advantage of this method is the recording can last more than 30 minutes in which we only get the whisker motion data and other methods like standard image capturing would exhaust the memory usage of the computer. The extracted data was smoothed in MATLAB and the power spectrum was calculated with a frequency resolution of 0.1 Hz.\\

\subsection*{PIV Set-up}
A 3 mm thick LED light sheet was focused on the region near the long whisker (Whisker 3) portion of the model. Neutrally buoyant particles of 50 microns were seeded in the water tunnel, and it is recirculated within the tunnel until we observe the right amount of particles required for the PIV. The images were captured with Phantom Miro M310 camera at a frame rate of 700 Hz and for a flow time of 4 seconds. Due to curvature and the orientation of the whiskers, the PIV was done in a plane so that two whiskers were on the PIV plane. The post-processing of PIV images was done inside Dantec Dynamic Studio software with adaptive PIV algorithm to calculate the velocity with a pair of two successive images and the velocity was averaged with the 3x3 filter. The Strouhal frequency of the upstream disturbance is calculated from the cross velocity spectrum of the wake, and the vortex core was visualised with an absolute 'Q' criterion method. 

\bibliography{library}

\begin{thebibliography}{10}
\urlstyle{rm}
\expandafter\ifx\csname url\endcsname\relax
  \def\url#1{\texttt{#1}}\fi
\expandafter\ifx\csname urlprefix\endcsname\relax\def\urlprefix{URL }\fi
\expandafter\ifx\csname doiprefix\endcsname\relax\def\doiprefix{DOI: }\fi
\providecommand{\bibinfo}[2]{#2}
\providecommand{\eprint}[2][]{\url{#2}}

\bibitem{ref1}
\bibinfo{author}{Bleckmann, H.} \& \bibinfo{author}{Zelick, R.}
\newblock \bibinfo{journal}{\bibinfo{title}{Lateral line system of fish}}.
\newblock {\emph{\JournalTitle{Integrative Zoology}}}
  \textbf{\bibinfo{volume}{4}}, \bibinfo{pages}{13--25} (\bibinfo{year}{2009}).

\bibitem{ref2}
\bibinfo{author}{Leitch, D.~B.} \& \bibinfo{author}{Catania, K.~C.}
\newblock \bibinfo{journal}{\bibinfo{title}{Structure, innervation and response
  properties of integumentary sensory organs in crocodilians}}.
\newblock {\emph{\JournalTitle{Journal of Experimental Biology}}}
  \textbf{\bibinfo{volume}{215}}, \bibinfo{pages}{4217--4230}
  (\bibinfo{year}{2012}).

\bibitem{ref3}
\bibinfo{author}{Cahlander, D.~A.}, \bibinfo{author}{McCue, J. J.~G.} \&
  \bibinfo{author}{Webster, F.~A.}
\newblock \bibinfo{journal}{\bibinfo{title}{The determination of distance by
  echolocating bats}}.
\newblock {\emph{\JournalTitle{Nature}}} \textbf{\bibinfo{volume}{201}},
  \bibinfo{pages}{544--546} (\bibinfo{year}{1964}).

\bibitem{ref4}
\bibinfo{author}{Ketten, D.~R.}
\newblock \emph{\bibinfo{title}{Cetacean Ears}}, \bibinfo{pages}{43--108}
  (\bibinfo{publisher}{Springer New York}, \bibinfo{address}{New York, NY},
  \bibinfo{year}{2000}).

\bibitem{Knight2489}
\bibinfo{author}{Knight, K.}
\newblock \bibinfo{journal}{\bibinfo{title}{Whisking whiskers tell rats about
  surroundings}}.
\newblock {\emph{\JournalTitle{Journal of Experimental Biology}}}
  \textbf{\bibinfo{volume}{218}}, \bibinfo{pages}{2489--2489}
  (\bibinfo{year}{2015}).

\bibitem{Dehnhardt1998}
\bibinfo{author}{Dehnhardt, G.}, \bibinfo{author}{Mauck, B.} \&
  \bibinfo{author}{Bleckmann, H.}
\newblock \bibinfo{journal}{\bibinfo{title}{Seal whiskers detect water
  movements}}.
\newblock {\emph{\JournalTitle{Nature}}} \textbf{\bibinfo{volume}{394}},
  \bibinfo{pages}{235--236} (\bibinfo{year}{1998}).

\bibitem{Dehnhardt102}
\bibinfo{author}{Dehnhardt, G.}, \bibinfo{author}{Mauck, B.},
  \bibinfo{author}{Hanke, W.} \& \bibinfo{author}{Bleckmann, H.}
\newblock \bibinfo{journal}{\bibinfo{title}{Hydrodynamic trail-following in
  harbor seals ({P}hoca vitulina)}}.
\newblock {\emph{\JournalTitle{Science}}} \textbf{\bibinfo{volume}{293}},
  \bibinfo{pages}{102--104} (\bibinfo{year}{2001}).

\bibitem{Glaser2011}
\bibinfo{author}{Gl{\"a}ser, N.}, \bibinfo{author}{Wieskotten, S.},
  \bibinfo{author}{Otter, C.}, \bibinfo{author}{Dehnhardt, G.} \&
  \bibinfo{author}{Hanke, W.}
\newblock \bibinfo{journal}{\bibinfo{title}{Hydrodynamic trail following in a
  california sea lion ({Z}alophus californianus)}}.
\newblock {\emph{\JournalTitle{Journal of Comparative Physiology A}}}
  \textbf{\bibinfo{volume}{197}}, \bibinfo{pages}{141--151}
  (\bibinfo{year}{2011}).

\bibitem{Dehnhardt}
\bibinfo{author}{Dehnhardt, G.} \& \bibinfo{author}{Mauck, B.}
\newblock chap. \bibinfo{chapter}{Mechanoreception in Secondarily Aquatic
  Vertebrates}, \bibinfo{pages}{295--314}.
\newblock Adaptations in Secondarily Aquatic Vertebrates
  (\bibinfo{publisher}{University of California Press}, \bibinfo{year}{2008}).

\bibitem{Hanke2665}
\bibinfo{author}{Hanke, W.} \emph{et~al.}
\newblock \bibinfo{journal}{\bibinfo{title}{Harbor seal vibrissa morphology
  suppresses vortex-induced vibrations}}.
\newblock {\emph{\JournalTitle{Journal of Experimental Biology}}}
  \textbf{\bibinfo{volume}{213}}, \bibinfo{pages}{2665--2672}
  (\bibinfo{year}{2010}).

\bibitem{Hanke2013}
\bibinfo{author}{Hanke, W.}, \bibinfo{author}{Wieskotten, S.},
  \bibinfo{author}{Marshall, C.} \& \bibinfo{author}{Dehnhardt, G.}
\newblock \bibinfo{journal}{\bibinfo{title}{Hydrodynamic perception in true
  seals ({P}hocidae) and eared seals ({O}tariidae)}}.
\newblock {\emph{\JournalTitle{Journal of Comparative Physiology A}}}
  \textbf{\bibinfo{volume}{199}}, \bibinfo{pages}{421--440}
  (\bibinfo{year}{2013}).

\bibitem{Miersch3077}
\bibinfo{author}{Miersch, L.} \emph{et~al.}
\newblock \bibinfo{journal}{\bibinfo{title}{Flow sensing by pinniped
  whiskers}}.
\newblock {\emph{\JournalTitle{Philosophical Transactions of the Royal Society
  of London B: Biological Sciences}}} \textbf{\bibinfo{volume}{366}},
  \bibinfo{pages}{3077--3084} (\bibinfo{year}{2011}).

\bibitem{beem_triantafyllou_2015}
\bibinfo{author}{Beem, H.~R.} \& \bibinfo{author}{Triantafyllou, M.~S.}
\newblock \bibinfo{journal}{\bibinfo{title}{Wake-induced ‘slaloming’
  response explains exquisite sensitivity of seal whisker-like sensors}}.
\newblock {\emph{\JournalTitle{Journal of Fluid Mechanics}}}
  \textbf{\bibinfo{volume}{783}}, \bibinfo{pages}{306–322}
  (\bibinfo{year}{2015}).

\bibitem{Schulte-Pelkum781}
\bibinfo{author}{Schulte-Pelkum, N.}, \bibinfo{author}{Wieskotten, S.},
  \bibinfo{author}{Hanke, W.}, \bibinfo{author}{Dehnhardt, G.} \&
  \bibinfo{author}{Mauck, B.}
\newblock \bibinfo{journal}{\bibinfo{title}{Tracking of biogenic hydrodynamic
  trails in harbour seals ({P}hoca vitulina)}}.
\newblock {\emph{\JournalTitle{Journal of Experimental Biology}}}
  \textbf{\bibinfo{volume}{210}}, \bibinfo{pages}{781--787}
  (\bibinfo{year}{2007}).

\bibitem{Wieskotten2194}
\bibinfo{author}{Wieskotten, S.}, \bibinfo{author}{Dehnhardt, G.},
  \bibinfo{author}{Mauck, B.}, \bibinfo{author}{Miersch, L.} \&
  \bibinfo{author}{Hanke, W.}
\newblock \bibinfo{journal}{\bibinfo{title}{Hydrodynamic determination of the
  moving direction of an artificial fin by a harbour seal ({P}hoca vitulina)}}.
\newblock {\emph{\JournalTitle{Journal of Experimental Biology}}}
  \textbf{\bibinfo{volume}{213}}, \bibinfo{pages}{2194--2200}
  (\bibinfo{year}{2010}).

\bibitem{Wieskotten1922}
\bibinfo{author}{Wieskotten, S.}, \bibinfo{author}{Mauck, B.},
  \bibinfo{author}{Miersch, L.}, \bibinfo{author}{Dehnhardt, G.} \&
  \bibinfo{author}{Hanke, W.}
\newblock \bibinfo{journal}{\bibinfo{title}{Hydrodynamic discrimination of
  wakes caused by objects of different size or shape in a harbour seal ({P}hoca
  vitulina)}}.
\newblock {\emph{\JournalTitle{Journal of Experimental Biology}}}
  \textbf{\bibinfo{volume}{214}}, \bibinfo{pages}{1922--1930}
  (\bibinfo{year}{2011}).

\bibitem{Kruger}
\bibinfo{author}{Kr{\"u}ger, Y.}, \bibinfo{author}{Hanke, W.},
  \bibinfo{author}{Miersch, L.} \& \bibinfo{author}{Dehnhardt, G.}
\newblock \bibinfo{journal}{\bibinfo{title}{Detection and direction
  discrimination of single vortex rings by harbour seals ({P}hoca vitulina)}}.
\newblock {\emph{\JournalTitle{Journal of Experimental Biology}}}
  \textbf{\bibinfo{volume}{221}} (\bibinfo{year}{2018}).

\bibitem{Niesterok2364}
\bibinfo{author}{Niesterok, B.}, \bibinfo{author}{Dehnhardt, G.} \&
  \bibinfo{author}{Hanke, W.}
\newblock \bibinfo{journal}{\bibinfo{title}{Hydrodynamic sensory threshold in
  harbour seals ({P}hoca vitulina) for artificial flatfish breathing
  currents}}.
\newblock {\emph{\JournalTitle{Journal of Experimental Biology}}}
  \textbf{\bibinfo{volume}{220}}, \bibinfo{pages}{2364--2371}
  (\bibinfo{year}{2017}).

\bibitem{Murphy2017}
\bibinfo{author}{Murphy, C.~T.}, \bibinfo{author}{Reichmuth, C.},
  \bibinfo{author}{Eberhardt, W.~C.}, \bibinfo{author}{Calhoun, B.~H.} \&
  \bibinfo{author}{Mann, D.~A.}
\newblock \bibinfo{journal}{\bibinfo{title}{Seal whiskers vibrate over broad
  frequencies during hydrodynamic tracking}}.
\newblock {\emph{\JournalTitle{Scientific Reports}}}
  \textbf{\bibinfo{volume}{7}}, \bibinfo{pages}{8350} (\bibinfo{year}{2017}).

\bibitem{Subramaniam2017}
\bibinfo{author}{Subramaniam, V.}, \bibinfo{author}{Alvarado, P. V.~y.} \&
  \bibinfo{author}{Weymouth, G.}
\newblock \emph{\bibinfo{title}{Sensing on Robots Inspired by Nature}},
  \bibinfo{pages}{77--110} (\bibinfo{publisher}{Springer International
  Publishing}, \bibinfo{address}{Cham}, \bibinfo{year}{2017}).

\bibitem{Solomon2006}
\bibinfo{author}{Solomon, J.~H.} \& \bibinfo{author}{Hartmann, M.~J.}
\newblock \bibinfo{journal}{\bibinfo{title}{Robotic whiskers used to sense
  features}}.
\newblock {\emph{\JournalTitle{Nature}}} \textbf{\bibinfo{volume}{443}},
  \bibinfo{pages}{525 EP --} (\bibinfo{year}{2006}).

\bibitem{William}
\bibinfo{author}{Eberhardt, W.~C.} \emph{et~al.}
\newblock \bibinfo{journal}{\bibinfo{title}{Development of an artificial sensor
  for hydrodynamic detection inspired by a seal’s whisker array}}.
\newblock {\emph{\JournalTitle{Bioinspiration \& Biomimetics}}}
  \textbf{\bibinfo{volume}{11}}, \bibinfo{pages}{056011}
  (\bibinfo{year}{2016}).

\bibitem{Nudds2244}
\bibinfo{author}{Nudds, R.~L.}, \bibinfo{author}{John, E.~L.},
  \bibinfo{author}{Keen, A.~N.} \& \bibinfo{author}{Shiels, H.~A.}
\newblock \bibinfo{journal}{\bibinfo{title}{Rainbow trout provide the first
  experimental evidence for adherence to a distinct strouhal number during
  animal oscillatory propulsion}}.
\newblock {\emph{\JournalTitle{Journal of Experimental Biology}}}
  \textbf{\bibinfo{volume}{217}}, \bibinfo{pages}{2244--2249}
  (\bibinfo{year}{2014}).

\bibitem{BAINBRIDGE109}
\bibinfo{author}{Bainbridge, R.}
\newblock \bibinfo{journal}{\bibinfo{title}{The speed of swimming of fish as
  related to size and to the frequency and amplitude of the tail beat}}.
\newblock {\emph{\JournalTitle{Journal of Experimental Biology}}}
  \textbf{\bibinfo{volume}{35}}, \bibinfo{pages}{109--133}
  (\bibinfo{year}{1958}).

\bibitem{Hanke1193}
\bibinfo{author}{Hanke, W.}, \bibinfo{author}{Brucker, C.} \&
  \bibinfo{author}{Bleckmann, H.}
\newblock \bibinfo{journal}{\bibinfo{title}{The ageing of the low-frequency
  water disturbances caused by swimming goldfish and its possible relevance to
  prey detection}}.
\newblock {\emph{\JournalTitle{Journal of Experimental Biology}}}
  \textbf{\bibinfo{volume}{203}}, \bibinfo{pages}{1193--1200}
  (\bibinfo{year}{2000}).

\bibitem{RJ2009}
\bibinfo{author}{RJ, S.}, \bibinfo{author}{B, A.} \& \bibinfo{author}{PS, H.}
\newblock \bibinfo{journal}{\bibinfo{title}{Seals, sandeels and salmon: diet of
  harbour seals in st. andrews bay and the tay estuary, southeast scotland}}.
\newblock {\emph{\JournalTitle{Marine Ecology Progress Series}}}
  \textbf{\bibinfo{volume}{390}}, \bibinfo{pages}{265--276}
  (\bibinfo{year}{2009}).

\bibitem{Nerve}
\bibinfo{author}{Dehnhardt, G.}, \bibinfo{author}{Mauck, B.} \&
  \bibinfo{author}{Hyv{\"a}rinen, H.}
\newblock \bibinfo{title}{The functional significance of the vibrissal system
  of marine mammals}.
\newblock In \bibinfo{editor}{Baumann, K.~I.}, \bibinfo{editor}{Halata, Z.} \&
  \bibinfo{editor}{Moll, I.} (eds.) \emph{\bibinfo{booktitle}{The Merkel
  Cell}}, \bibinfo{pages}{127--135} (\bibinfo{publisher}{Springer Berlin
  Heidelberg}, \bibinfo{address}{Berlin, Heidelberg}, \bibinfo{year}{2003}).

\bibitem{Sawyer}
\bibinfo{author}{Sawyer, E.~K.}, \bibinfo{author}{Turner, E.~C.} \&
  \bibinfo{author}{Kaas, J.~H.}
\newblock \bibinfo{journal}{\bibinfo{title}{Somatosensory brainstem, thalamus,
  and cortex of the california sea lion ({Z}alophus californianus)}}.
\newblock {\emph{\JournalTitle{Journal of Comparative Neurology}}}
  \textbf{\bibinfo{volume}{524}}, \bibinfo{pages}{1957--1975}.

\bibitem{VIV}
\bibinfo{author}{Williamson, C.} \& \bibinfo{author}{Govardhan, R.}
\newblock \bibinfo{journal}{\bibinfo{title}{Vortex-induced vibrations}}.
\newblock {\emph{\JournalTitle{Annual Review of Fluid Mechanics}}}
  \textbf{\bibinfo{volume}{36}}, \bibinfo{pages}{413--455}
  (\bibinfo{year}{2004}).

\bibitem{Cylstrouhal}
\bibinfo{author}{Williamson, C. H.~K.}
\newblock \bibinfo{journal}{\bibinfo{title}{Vortex dynamics in the cylinder
  wake}}.
\newblock {\emph{\JournalTitle{Annual Review of Fluid Mechanics}}}
  \textbf{\bibinfo{volume}{28}}, \bibinfo{pages}{477--539}
  (\bibinfo{year}{1996}).

\bibitem{Strouhalref}
\bibinfo{author}{Fey, U.}, \bibinfo{author}{König, M.} \&
  \bibinfo{author}{Eckelmann, H.}
\newblock \bibinfo{journal}{\bibinfo{title}{A new strouhal–reynolds-number
  relationship for the circular cylinder in the range $47 \leq {R}e \leq
  2{X}10^5$}}.
\newblock {\emph{\JournalTitle{Physics of Fluids}}}
  \textbf{\bibinfo{volume}{10}}, \bibinfo{pages}{1547--1549}
  (\bibinfo{year}{1998}).

\bibitem{Muthu}
\bibinfo{author}{Muthuramalingam, M.} \& \bibinfo{author}{Br\"{u}cker, C.}
\newblock \bibinfo{journal}{\bibinfo{title}{On the interaction of von-karman
  vortices on whisker-like elements}}.
\newblock {\emph{\JournalTitle{Unpublished results}}}  (\bibinfo{year}{2018}).

\bibitem{MONTGOMERY1994145}
\bibinfo{author}{Montgomery, J.~C.} \& \bibinfo{author}{Bodznick, D.}
\newblock \bibinfo{journal}{\bibinfo{title}{An adaptive filter that cancels
  self-induced noise in the electrosensory and lateral line mechanosensory
  systems of fish}}.
\newblock {\emph{\JournalTitle{Neuroscience Letters}}}
  \textbf{\bibinfo{volume}{174}}, \bibinfo{pages}{145 -- 148}
  (\bibinfo{year}{1994}).

\bibitem{jeong_hussain_1995}
\bibinfo{author}{Jeong, J.} \& \bibinfo{author}{Hussain, F.}
\newblock \bibinfo{journal}{\bibinfo{title}{On the identification of a
  vortex}}.
\newblock {\emph{\JournalTitle{Journal of Fluid Mechanics}}}
  \textbf{\bibinfo{volume}{285}}, \bibinfo{pages}{69–94}
  (\bibinfo{year}{1995}).

\bibitem{Green1995}
\bibinfo{author}{Green, S.~I.}
\newblock \emph{\bibinfo{title}{Introduction to Vorticity}},
  \bibinfo{pages}{1--34} (\bibinfo{publisher}{Springer Netherlands},
  \bibinfo{address}{Dordrecht}, \bibinfo{year}{1995}).

\bibitem{Oladazimi2018}
\bibinfo{author}{Oladazimi, M.}, \bibinfo{author}{Brendel, W.} \&
  \bibinfo{author}{Schwarz, C.}
\newblock \bibinfo{journal}{\bibinfo{title}{Biomechanical texture coding in rat
  whiskers}}.
\newblock {\emph{\JournalTitle{Scientific Reports}}}
  \textbf{\bibinfo{volume}{8}}, \bibinfo{pages}{11139} (\bibinfo{year}{2018}).

\bibitem{diameter}
\bibinfo{author}{Ginter~Summarell, C.~C.}, \bibinfo{author}{Ingole, S.},
  \bibinfo{author}{Fish, F.~E.} \& \bibinfo{author}{Marshall, C.~D.}
\newblock \bibinfo{journal}{\bibinfo{title}{Comparative analysis of the
  flexural stiffness of pinniped vibrissae}}.
\newblock {\emph{\JournalTitle{PLOS ONE}}} \textbf{\bibinfo{volume}{10}},
  \bibinfo{pages}{1--15} (\bibinfo{year}{2015}).

\bibitem{LEALJUNIOR2018106}
\bibinfo{author}{Leal-Junior, A.}, \bibinfo{author}{Frizera, A.},
  \bibinfo{author}{Marques, C.} \& \bibinfo{author}{Pontes, M.~J.}
\newblock \bibinfo{journal}{\bibinfo{title}{Mechanical properties
  characterization of polymethyl methacrylate polymer optical fibers after
  thermal and chemical treatments}}.
\newblock {\emph{\JournalTitle{Optical Fiber Technology}}}
  \textbf{\bibinfo{volume}{43}}, \bibinfo{pages}{106 -- 111}
  (\bibinfo{year}{2018}).

\end{thebibliography}

\section*{Acknowledgements}

The position of Professor Christoph Br\"{u}cker is co-funded by BAE SYSTEMS and the Royal Academy of Engineering (Research Chair No. RCSRF1617$\backslash$4$\backslash$11, which is gratefully acknowledged. The position of MSc Muthukumar Muthuramalingam was funded by the Deutsche Forschungsgemeinschaft in the DFG project BR 1494/32-1, which largely supported the work described herein.The high speed camera was funded by the Deutsche Forschungsgemeinschaft in the DFG project BR 1494/30-1. The CT scan data for the modelling was provided by Massimo Bernardi, Department of Geology and Palaeontology, MUSE - Museo delle Scienze, Trento, Italy. Acknowledgments for Prof. Nicola Pugno, Department of Civil, Environmental and Mechanical Engineering, University of Trento, Italy.
Comments from Prof. Horst Bleckmann, University of Bonn, Germany is gratefully acknowledged. Edward Talboys from City University of London is acknowledged for the manuscript corrections. 

\section*{Author contributions statement}

C.B. (Professor) was in charge of collecting the scanned model of sea lion and principal Investigator of the project. The experimental setup, experiments, post processing of data was done by M. M who is a PhD Student at City University of London. The manuscript was based on the ideas and discussion on the hydrodynamic sensors.

\section*{Additional information}

\textbf{Accession codes} (where applicable);\\ \textbf{Competing interests} The authors declare no competing interests..

\end{document}